\documentclass{article}
\usepackage{spconf,amsmath,graphicx}
\usepackage{MnSymbol}
\usepackage{amsthm}
\usepackage{enumitem}
\usepackage{amsfonts}
\usepackage[font=small,skip=.5pt]{caption}

\usepackage[tight,footnotesize]{subfigure}
\usepackage{multirow}
\usepackage{pbox}
\usepackage{url}
\usepackage{comment}

\usepackage{bm}

\long\def\comment#1{}

\newfont{\bbb}{msbm10 scaled 700}

\newfont{\bb}{msbm10 scaled 1100}

\newcommand{\ev}{{\bf e}}

\newcommand{\pv}{{\bf p}}

\newcommand{\rv}{{\bf r}}

\newcommand{\uv}{{\bf u}}

\newcommand{\xv}{{\bf x}}
\newcommand{\yv}{{\bf y}}

\newcommand{\zerov}{{\bf 0}}
\newcommand{\onev}{{\bf 1}}

\newcommand{\Am}{{\bf A}}
\newcommand{\Bm}{{\bf B}}

\newcommand{\Dm}{{\bf D}}

\newcommand{\Gm}{{\bf G}}

\newcommand{\Id}{{\bf I}}

\newcommand{\Km}{{\bf K}}
\newcommand{\Lm}{{\bf L}}

\newcommand{\Sm}{{\bf S}}

\newcommand{\Um}{{\bf U}}
\newcommand{\Wm}{{\bf W}}
\newcommand{\Vm}{{\bf V}}

\newcommand{\Ec}{{\cal E}}

\newcommand{\Gc}{{\cal G}}

\newcommand{\Jc}{{\cal J}}

\newcommand{\Lc}{{\cal L}}

\newcommand{\Nc}{{\cal N}}

\newcommand{\Qc}{{\cal Q}}

\newcommand{\Tc}{{\cal T}}
\newcommand{\Uc}{{\cal U}}

\newcommand{\Vc}{{\cal V}}

\newcommand{\xiv}{\hbox{\boldmath$\xi$}}

\newcommand{\Lambdam}{\hbox{\boldmath$\Lambda$}}

\newcommand{\Xim}{\hbox{\boldmath$\Xi$}}

\renewcommand{\det}{{\hbox{det}}}

\newtheorem{theorem}{Theorem}

\newtheorem{lemma}{Lemma}

\title{Closed Form Solutions of Combinatorial Graph Laplacian Estimation Under Acyclic Topology Constraints}
\name{Keng-Shih Lu and Antonio Ortega
}
\address{Department of Electrical Engineering, 
University of Southern California, Los Angeles, USA}

\begin{document}
\ninept
\maketitle
\begin{abstract}
How to obtain a graph from data samples is an important problem in graph signal processing. One way to formulate this graph learning problem is based on Gaussian maximum likelihood estimation, possibly under particular topology constraints. To solve this problem, we typically require iterative convex optimization solvers. In this paper, we show that when the target graph topology does not contain any cycle, then the solution has a closed form in terms of the empirical covariance matrix. This enables us to efficiently construct a tree graph from data, even if there is only a single data sample available. We also provide an error bound of the objective function when we use the same solution to approximate a cyclic graph. 
As an example, we consider an image denoising problem, in which for each input image we construct a graph based on the theoretical result. We then apply low-pass graph filters based on this graph. Experimental results show that the weights given by the graph learning solution lead to better denoising results than the bilateral weights under some conditions. 
%
%

\end{abstract}
\section{Introduction}
\label{sec:intro}
Graph signal processing \cite{Shuman2013,Sandryhaila2013} extends conventional signal processing tools to signals supported on graphs, in which the vertices and edges are used to model objects and their pairwise relations. This framework allows us to apply signal processing techniques while taking into consideration the connectivity/correlation among vertices, providing more flexibility. Because of this convenience, graph signal processing has found applications in sensor networks \cite{Jablonski2017}, semi-supervised and unsupervised learning \cite{Gadde2014}, social networks \cite{Hassan2016}, image and video processing \cite{Shen2010}, and so on.

Graph learning \cite{Dong2016,Egilmez2017} is an important problem in graph signal processing because the there are scenarios where the best graph to process data is not known ahead of time. Graph learning typically involves two stages: topology inference and weight estimation. Topology inference aims to identify which pairs of nodes should be connected. In some scenarios, the graphs are required to have certain structures, such as tree (acyclic) structure \cite{Shen2009} or bipartite structure \cite{Narang2012}, and solving the optimal topology can involve solving NP-hard combinatorial problems \cite{Pavez2017}. For weight estimation, the goal is to assign weight values to the edges with a given graph topology. In particular, given the empirical covariance matrix $\Sm=\sum_{i=1}^N \xv_i\xv_i^T/N$ from $N$ data samples $\xv_i$, the combinatorial graph Laplacian (CGL) estimation problem can be formulated as 
\begin{equation}
\label{eq:glpest}
  \underset{\Lm\in\mathbb{L}(\Ec)}{\text{minimize}} \quad -\log|\Lm|_\dagger
    +\text{trace}(\Lm\Sm)+\alpha\|\Lm\|_{\text{1,off}},
\end{equation}
where $\mathbb{L}(\Ec)$ is the set of CGLs with edge set $\Ec$, and $\|\Lm\|_{\text{1,off}}$ is the absolute sum of all off-diagonal elements of $\Lm$. Note that the pseudo-determinant $|\Lm|_\dagger$ is required here because CGL matrices are singular. The formulation \eqref{eq:glpest} is a maximum a posteriori (MAP) parameter estimation of a Gaussian Markov Random Field (GMRF) $\xv\sim\Nc(\zerov,\Sigma=\Lm^\dagger)$. In \cite{Egilmez2017}, iterative methods based on block-coordinate descent algorithms were proposed and shown to provide better efficiency than other existing methods. 

%
%
In this work, we focus on weight estimation given a graph topology, and seek to find a fast solution beyond the iterative algorithm proposed in \cite{Egilmez2017} under some specific graph topology conditions. Specifically, we aim to provide answers to the following questions. 1)~Can we find sufficient conditions on the graph topology for the  optimal weights to have a closed form expression in terms of the covariance matrix $\Sm$ or the data samples $\xv_i$? 2)~If such solution exists, can we design a graph construction method based on it, for applications where an iterative solution is too complex? The contributions of this work are summarized as follows. Firstly, we show that if the given graph topology has an acyclic structure, the solution of \eqref{eq:glpest} has a closed form. This allows us to construct the graph associated to the optimal GMRF estimate from a small number of  data samples (including from just a single data sample). Secondly, we provide an error bound for the objective function in the case when we apply this method to graphs with cycles. Finally, we illustrate the potential benefits of our approach with  experimental results in image denoising. 

In \cite{Fattahi2017}, it is shown that the graphical Lasso problem \cite{Friedman2008} admits a closed form solution when the soft thresholded sample correlation matrix has an acyclic structure. While this result and its corresponding conditions are somewhat similar to our results, there are several essential differences between those two problems. Firstly, the problem studied in \cite{Fattahi2017} takes the {\it correlation} matrix as input while typical graph estimation problem uses the {\it covariance} matrix; as a result, the closed form provided in this paper is simpler in terms of data samples, providing a convenient approach for graph weight construction from data. Secondly, the graph Laplacian estimation problem we consider here incorporates additional constraints in addition to those encountered in the graphical Lasso problem, so that a solution for our problem cannot be easily derived the graphical Lasso solution. To the best of our knowledge, closed form solutions for problem of \eqref{eq:glpest} have not been presented in the literature.

The rest of this paper is organized as follows. In Section \ref{sec:preliminaries} we introduce some background knowledge in graph signal processing and formulate the problem. In Section \ref{sec:glpest} we show the main theoretical results for graph weight estimation under the acyclic topology constraint. We also present a case study for line graphs and provide an error bound for our solution when applied to cyclic graphs. In Section \ref{sec:experiments} we show some experimental results on synthetic data and real images. Finally conclusions are presented in Section \ref{sec:conclusion}.

\section{Preliminaries}
\label{sec:preliminaries}
We consider a weighted undirected graph $\Gc=(\Vc,\Ec,\Wm)$ with the nodes representing attributes of a data sample (e.g, pixels in an image block), and edges describing the relations between attributes (e.g, pairwise pixel correlation). We let $|\Vc|=n$ and $|\Ec|=m$. Each element $w_{i,j}$ in $\Wm$ is the weight for an edge $(i,j)\in\Ec$, and $w_{i,j}=0$ if $(i,j)\notin\Ec$. The CGL matrix is defined as $\Dm-\Wm$, where $\Dm$ is the diagonal degree matrix with $d_{i,i}=\sum_{j=1}^n w_{i,j}$ and $d_{i,j}=0,\; i\neq j$. 

The Graph Fourier Transform (GFT) coefficients are projections of graph signal $\xv$ onto each eigenvector of the graph Laplacian matrix. That is, with eigendecomposition $\Lm=\Um\Lambdam\Um^T$, the GFT vector of $\xv$ is $\hat{\xv}=\Um^T\xv$. With GFT, we can define graph filters as $\yv=\Um h(\Lambdam)\Um^T\xv$ with $h(\lambda)$ the spectral response of the filter. 

It is shown in \cite{Egilmez2017} that the CGL problem \eqref{eq:glpest} has the following equivalent objective function.
\begin{equation}
\label{eq:objfunc_cgl}
  \Jc(\Lm) := -\text{log det}(\Lm+\onev\onev^T/n)+\text{trace}(\Lm\Km),
\end{equation}
where $\Km=\Sm+\alpha(\Id-\onev\onev^T)+\onev\onev^T/n$. 
With a given graph topology $\Ec$, we can formulate \eqref{eq:glpest} more explicitly as
\begin{align}
\label{eq:prob_cgl}
  & \underset{\Lm}{\text{minimize}} \quad \Jc(\Lm) \nonumber\\
  & \text{subject to} \quad \Lm\succeq 0, \quad \Lm\onev=\zerov, \nonumber\\
  & \hspace{1.6cm} \ell_{i,j}=0,\;\text{if }(i,j)\notin\Ec, \nonumber\\
  & \hspace{1.6cm} \ell_{i,j}<0,\;\text{if }(i,j)\in\Ec.
\end{align}

\section{Acyclic Graph Weight Estimation}
\label{sec:glpest}
In this section, we present the proof to our main result in Section \ref{sec:represent} and \ref{sec:cf}, discuss a special case of tree graphs--line graphs in Section \ref{sec:line}, and provide an error bound for applying the solution to graphs with cycles in Section \ref{sec:erbd}.

\subsection{Representation of CGL with the Incidence Matrix}
\label{sec:represent}
Consider the oriented incidence matrix of a graph: $\Xim=(\xiv_1,\dots,\xiv_m)\in\mathbb{R}^{n\times m}$. Each of its columns $\xiv_j$  has two nonzero elements, 1 and -1,  representing an edge $\varepsilon_j$ connecting nodes $s_j$ and $t_j$; that is, $\xiv_j=\ev_{s_j}-\ev_{t_j}$ for $\varepsilon_j=(s_j,t_j)\in\Ec$. We can express $\Lm$ in terms of the edge weights and $\Xim$:
\begin{align}
\label{eq:l2w}
  \Lm &= \Dm-\Wm \nonumber\\ 
  &=\sum_{j=1}^m \underbrace{w_{s_j,t_j}(\ev_{s_j}\ev_{s_j}^T+\ev_{t_j}\ev_{t_j}^T)}_{\text{contribution of $\varepsilon_j$ to $\Dm$}} - \underbrace{w_{s_j,t_j}(\ev_{s_j}\ev_{t_j}^T + \ev_{t_j}\ev_{s_j}^T)}_{\text{contribution of $\varepsilon_j$ to $\Wm$}} \nonumber\\
  &= \sum_{j=1}^m w_{s_j,t_j}\xiv_j\xiv_j^T \nonumber\\
  &= \Xim\;\text{diag}(w_{s_1,t_1},\dots,w_{s_m,t_m})\;\Xim^T.
\end{align}
Note that, as $\varepsilon_j$ is undirected, either $\xiv_j=\ev_{s_j}-\ev_{t_j}$ or $\xiv_j=\ev_{t_j}-\ev_{s_j}$ is allowed; equation \eqref{eq:l2w} holds either way, and the choice does not affect  the following results. From now on, for simplicity, we define $u_j:=w_{s_j,t_j}$ to represent weights with a single index, and also define a weight vector as $\uv=(u_1,\dots,u_m)^T$ for compact notation.

To change the variable $\Lm$ in \eqref{eq:objfunc_cgl} to $\uv$, we use \eqref{eq:l2w} and get
\begin{equation}
\label{eq:l_cgl}
  \Lm+\onev\onev^T/n
  =\sum_{j=1}^m u_j\xiv_j\xiv_j^T + (1/n)\onev\onev^T
  = \Gm\;\text{diag}(\uv^+)\;\Gm^T,
\end{equation}
where we define $\Gm:=(\Xim, \onev)=(\xiv_1,\dots,\xiv_m,\onev)$ and $\uv^+:=(u_1,\dots,u_m,1/n)^T$. From \eqref{eq:l2w}, we can regard $\Lm$ as an affine function of $\uv$. Since the composition of a convex function and an affine mapping is convex, the convexity of $\Jc(\Lm)$ in $\Lm$ implies that 
\[
  \Jc(\uv) = -\text{log det}\left[\Gm\;\text{diag}(\uv^+)\;\Gm^T\right] + \text{trace}(\Xim\;\text{diag}(\uv)\;\Xim^T\Km)
\]
is convex in $\uv$. 

One convenience provided by this representation $\Jc(\uv)$ is that the list of constraints in \eqref{eq:prob_cgl} can be reduced to a single constraint $\uv>\zerov$ because 1)~the topology constraints described by $\Ec$ are now encoded in $\Xim$, 2)~the constraint $\uv>\zerov$ implies $\Lm\succeq 0$, and 3)~$\Lm\onev=\zerov$ follows from $\xiv_j^T\onev=0$ for all $j$. The problem \eqref{eq:prob_cgl} now reads
\[
  \underset{\uv>\zerov}{\text{minimize}} \quad \Jc(\uv)
\]

\subsection{Closed Form Solution}
\label{sec:cf}
The derivative of $\Jc(\uv)$ with respect to  $u_j$ is
\begin{equation}
\label{eq:derivative}
  \frac{\partial \Jc(\uv)}{\partial u_j} = -\xiv_j^T\left[\Gm\;\text{diag}(\uv^+)\;\Gm^T\right]^{-1}\xiv_j + \xiv_j^T\Km\xiv_j.
\end{equation}
We are interested in the conditions of $\Gm$ for \eqref{eq:derivative} to have a closed form expression (without matrix inversion) in terms of $u_j$, so that we can solve the optimal $u_j$ by setting \eqref{eq:derivative} to zero. One hope is to consider the case with $m=n-1$ so that $\Gm$ is a square matrix. For the invertibility of $\Gm$, we consider the following lemma. 
\begin{lemma}[\cite{West2000}]
  The oriented incidence matrix of a graph with $n$ nodes and $k$ connected components has rank $n-k$. 
\end{lemma}
This means that when $m=n-1$, $\Xim$ has full column rank $n-1$ if and only if the graph is connected, or equivalently, acyclic. In this case, since $\onev$ is orthogonal to every column $\xiv_j$ of $\Xim$, the rank of $\Gm=(\Xim,\onev)$ is $n$. Thus, $\Gm$ is invertible, and \eqref{eq:derivative} can be expressed as 
\begin{equation}
\label{eq:derivative2}
  \frac{\partial \Jc(\uv)}{\partial u_j} = -\xiv_j^T\Gm^{-T}\;\text{diag}(\uv^+)^{-1}\Gm^{-1}\xiv_j + \xiv_j^T\Km\xiv_j.
\end{equation}
By definition of the inverse matrix and the fact that $\xiv_j$ is the $j$-th column of $\Gm$, we have $\Gm^{-1}\xiv_j=\ev_j$. Plugging it into \eqref{eq:derivative2} and setting \eqref{eq:derivative2} to zero, we have $u_j = 1/\left(\xiv_j^T\Km\xiv_j\right)$. 
Note that, with the expression $\Km=\Sm+\alpha(\Id-\onev\onev^T)+\onev\onev^T/n$ and $\Sm=\sum_{i=1}^N\xv_i\xv_i^T/N$, the weights can be expressed in terms of $\xv_i$:
\begin{equation}
\label{eq:cf1}
  w_{s,t} = \left[\frac{1}{N}\sum_{i=1}^N\left(\xv_i(s)-\xv_i(t)\right)^2+2\alpha\right]^{-1},\quad \text{for }(s,t)\in\Ec,
\end{equation}
It is clear that this value is always positive given $\alpha>0$, so $\uv>\zerov$ is satisfied, meaning that the derived form is indeed the closed form solution of \eqref{eq:prob_cgl}. This main result is summarized as follows.
\begin{theorem}
\label{thm:main}
If $\Ec$ corresponds to a connected, loopless, acyclic graph, then the optimal CGL solution for \eqref{eq:prob_cgl} is given in \eqref{eq:cf1}.
\end{theorem}
Omitting a rather similar proof, we also state the counterpart result for generalized graph Laplacian (GGL), where self-loops are allowed in the graph. 
\begin{theorem}
\label{thm:main_ggl}
If $\Ec$ corresponds to a connected, acyclic graph with one self-loop $v_k$, then the optimal GGL solution has edge weights given in \eqref{eq:cf1}, and the self-loop weight given by $v_k=1/(\ev_k^T\Km\ev_k)$, or
\begin{equation}
\label{eq:cf2}
  v_k = \left[\frac{1}{N}\sum_{i=1}^N\xv_i(k)^2\right]^{-1}
\end{equation}
\end{theorem}

\subsection{Case Study: Line Graphs}
\label{sec:line}
Line graphs are special cases of tree graphs. Their Laplacian matrices correspond to precision matrices of 1D first order GMRFs, where each interior node is only connected to its two immediate neighbors. In image and video compression, such GMRFs are used for modeling pixels and interpreting their spatial correlations \cite{Zhang2013}. Transforms derived from such GMRFs are shown to provide a compression gain as compared to the traditional 2D discrete cosine transform (DCT) \cite{Egilmez2015,Pavez2017icip}. 

\begin{figure}
\centering
\subfigure[Line graph, for columns]{
\includegraphics[width=.35\textwidth]{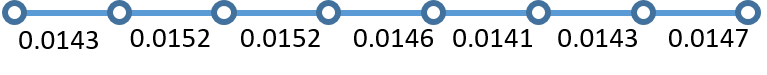}}
\subfigure[Symmetric line graph, for column]{
\includegraphics[width=.35\textwidth]{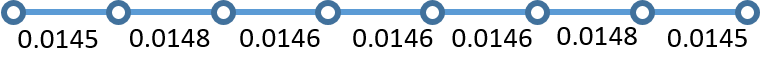}}
\subfigure[Line graph, for rows]{
\includegraphics[width=.35\textwidth]{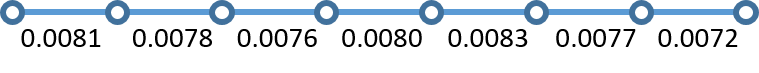}}
\subfigure[Symmetric line graph, for rows]{
\includegraphics[width=.35\textwidth]{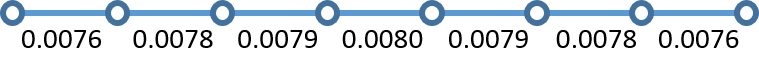}}
\caption{Line graphs learned from the lena image using the closed form construction with $\alpha=0$.}
\label{fig:lgs_lena}
\vspace{-.3cm}
\end{figure}

Since line graphs are acyclic, the solutions \eqref{eq:cf1} and \eqref{eq:cf2} can be used to learn line graphs (with at most one self-loop) from block samples. It means that loopless line graph learning for separable transforms can be achieved efficiently with the following steps: 
\begin{itemize}[noitemsep,topsep=0pt,parsep=0pt,partopsep=0pt]
\item[{\bf 1}] Collect 1D row or column samples $\rv_1,\dots,\rv_N$ (each with length $n$). 
\item[{\bf 2}] Compute mean square difference values: for $k=1,\dots,n-1$,
\[
  d_k = \frac{1}{N}\sum_{i=1}^N(\rv_i(k)-\rv_i(k+1))^2
\]
\item[{\bf 3}] Compute weights as $w_{k,k+1}=(d_k+2\alpha)^{-1}$.
\end{itemize}

When we constrain some weights in the line graph to be equal, the above procedure, with proper modifications, can still apply. In \cite{Lu2016}, the goal is to learn line graphs that are symmetric around the middle, which gives rise to a computation speedup using a butterfly stage. Following the steps from \eqref{eq:l_cgl} to \eqref{eq:derivative} with $u_{n-j-1}$ replaced by $u_j$, we also have a closed form solution, based on which we obtain the desired symmetric weights by adding one step between steps 2 and 3:
\begin{itemize}[noitemsep,topsep=0pt,parsep=0pt,partopsep=0pt]
\item[{\bf 2*}] Update $d_k$ as $d_k\leftarrow(d_k+d_{n-k-1})/2$.
\end{itemize}
As an example, we show in Fig. \ref{fig:lgs_lena} the loopless line graphs obtained from the above procedure from $8\times 8$ blocks of the grayscale Lena image, with and without symmetry constraints. 

\subsection{Error Bound for Cyclic Graphs}
\label{sec:erbd}
Theorems \ref{thm:main} and \ref{thm:main_ggl} apply to graphs with a very specific topology (i.e., acyclic graphs). Thus, we would be interested in using \eqref{eq:cf1} in order to estimate weights for graphs with more general topologies. We next provide 
a bound on the penalty with respect to the optimal solution when using the closed form solution of \eqref{eq:cf1} for a graph with a {\it general} connected topology (i.e., not cyclic). 

\begin{theorem}
\label{thm:erbd}
Let the number of nodes be $n$ and let a given connected graph topology be $\Ec=\{\varepsilon_1,\dots,\varepsilon_m\}$. Define $\Lm^*$ as the optimal solution of \eqref{eq:prob_cgl} and $\Lm$ as the solution constructed using \eqref{eq:cf1}. Then, 
\begin{equation*}
  0\leq \Jc(\Lm)-\Jc(\Lm^*)
  \leq \sum_{j=n}^m\left(1-\frac{u_j^*}{u_j}\right)+(m-n+1)\log\left(\frac{\max_i\; r_i}{\min_i\; r_i}\right),
\end{equation*}
where, with $i=1,\dots,m$,
\[
  r_i = \left\{\begin{array}{ll} u_i^*/u_i, & u_i^*>0 \\
  1, & u_i^*=0\end{array}\right.
\]
\end{theorem}
%
%
According to \eqref{eq:cf1}, $u_i$ are always positive, so this bound is always finite. The proof of this theorem will be included in our upcoming submission to a journal paper.


\section{Experiments}
\label{sec:experiments}

\subsection{Synthetic Data}
\begin{figure}
\centering
\subfigure[]{
\includegraphics[width=.24\textwidth]{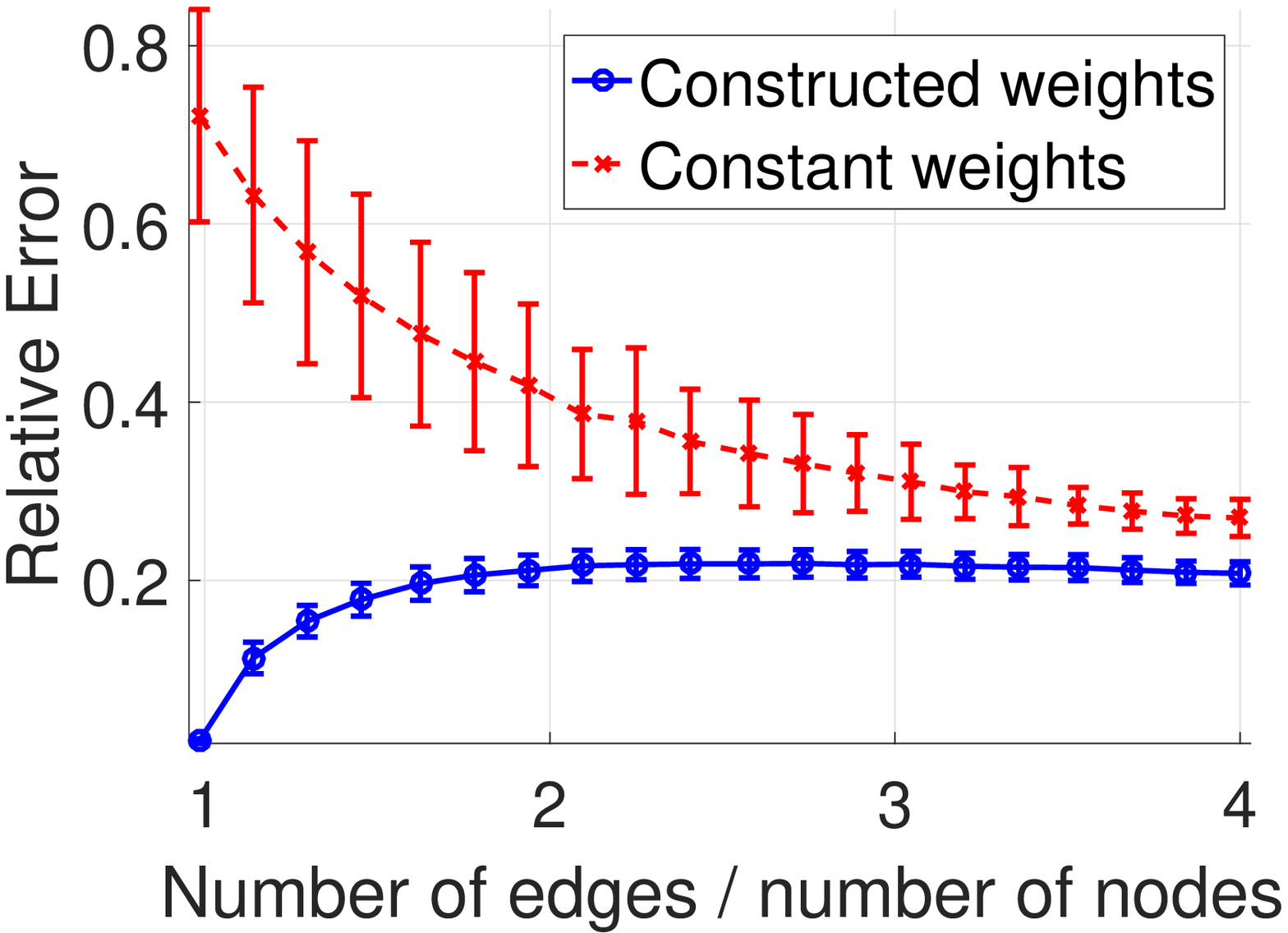}}%
\subfigure[]{
\includegraphics[width=.24\textwidth]{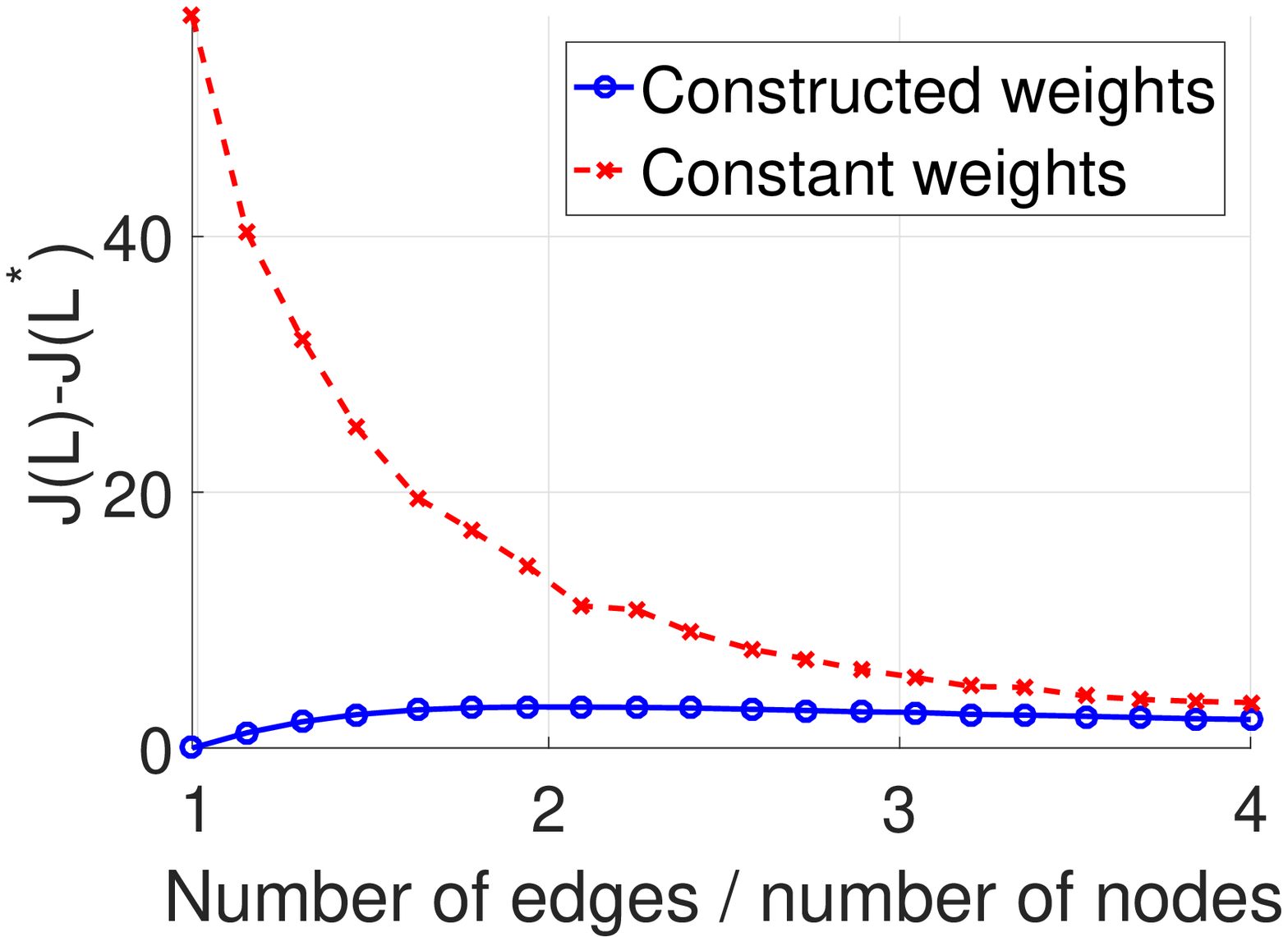}}
\caption{(a) Mean and standard deviation of relative errors between the optimal and constructed solutions. (b) Average difference between objective function values of optimal and constructed solutions. }
\label{fig:error}
\vspace{-.3cm}
\end{figure}

We first test the weight construction of \eqref{eq:cf1} on randomly generated connected graphs. We use $n=64$ nodes, and construct 500 connected graphs with a given number $m\geq n-1$ of edges. Ground truth weights of those edges are drawn from a uniform distribution $\Uc(0,1)$. For each graph $\Lm$, we generate $N=3200$ realizations $\xv_1,\dots,\xv_N$ of graph signals from the associated GMRF, $\Nc(\zerov,\Lm^\dagger)$. Then, we construct the graph based on \eqref{eq:cf1} with $\alpha=0$ using those $N$ realizations, and apply a scaling for normalization. The error between the constructed solution $\Lm_\text{con}$ and the true solution $\Lm^*$, is measured using the relative error (RE) in Frobenius norm: 
\[ \text{RE} = \| \Lm^* - \Lm_{\text{con}}\|_F/\| \Lm^* \|_F
\]
The same procedure is repeated for 20 different values of $m$. We plot the average of RE, and of $\Jc(\Lm_\text{con})-\Jc(\Lm^*)$ in Fig. \ref{fig:error}, versus $m/n$. We also plot the resulting errors with a constant weight construction, where all weights are assigned to be 0.5. 

In addition to the theoretical bound in Theorem \ref{thm:erbd}, which may be rather loose for practical cases. This experiment provides an empirical error measure, from which we observe \eqref{eq:cf1} is a reasonable approximation that yields much smaller error compared to constant weights, especially when the graph is sparse. 

\subsection{Edge-Adaptive Image Denoising}

\begin{figure}
\centering
\subfigure[]{
\includegraphics[width=.24\textwidth]{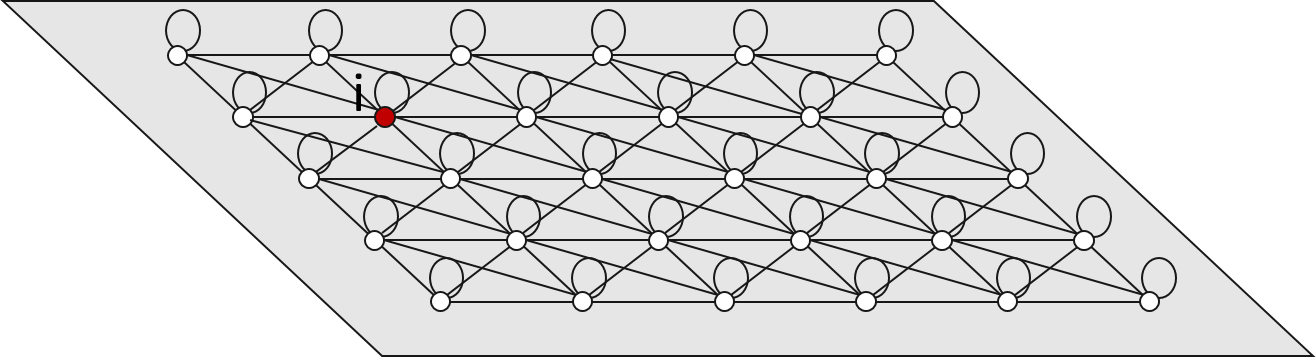}}%
\subfigure[]{
\includegraphics[width=.24\textwidth]{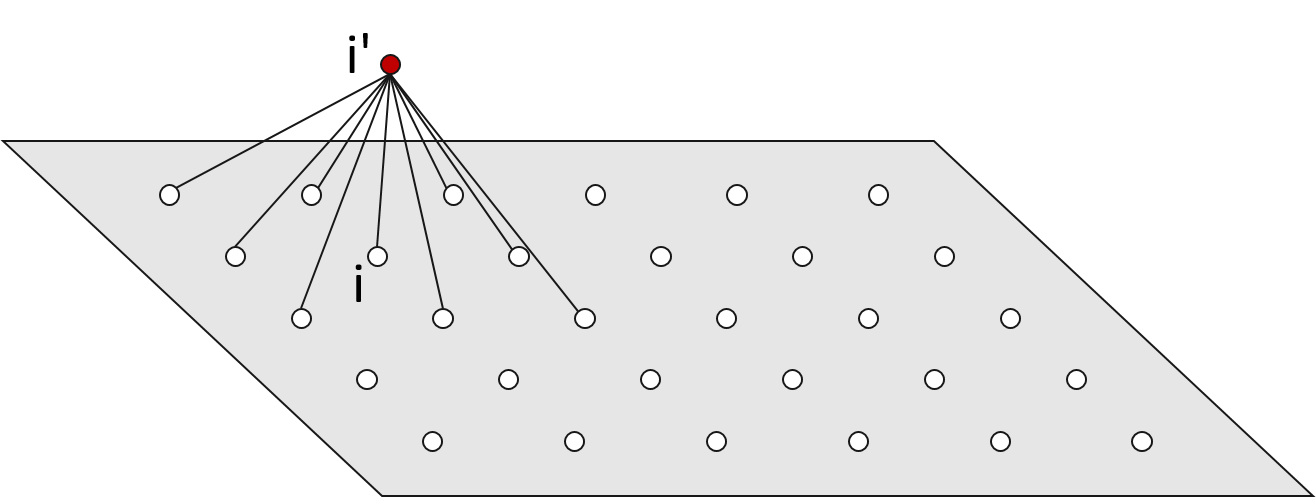}}
\caption{Illustration of the bilateral filter with $k=3$ regarded as a graph filter. The bilateral filter output at pixel $i$ is the graph filter output of node $i$ based on graph (a). It is, equivalently, also the graph filter output of node $i'$ based on the graph in (b).}
\label{fig:bflocal}
\end{figure}
\begin{figure}
\centering
\includegraphics[width=.24\textwidth]{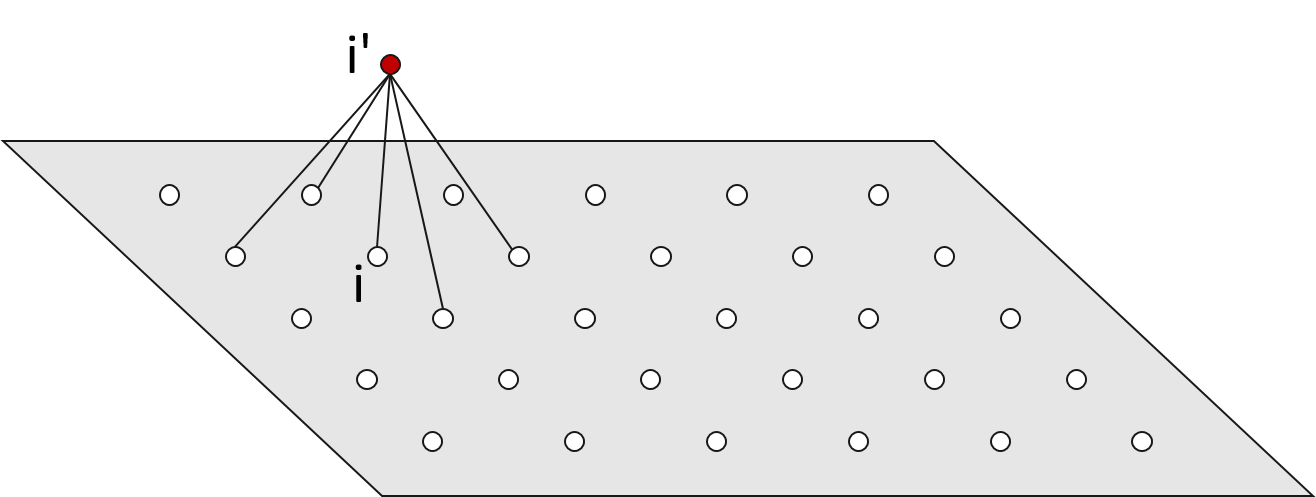}
\vspace{.1cm}
\caption{The local topology of the 5-neighbor graph considered in the experiment.}
\label{fig:bflocal_4conn}
\end{figure}

We consider image denoising as an application for the weight construction \eqref{eq:cf1}. The bilateral filter \cite{Tomasi1998} is a classical edge-aware image denoising filter, which takes the noisy image $\xv$ as input, then smooths each pixel $i$ using a weighted average of pixels $j$ in the $k\times k$ window around it:
\begin{equation}
\label{eq:bf_output}
  y_i=\sum_j\frac{w_{i,j}}{\sum_i w_{i,j}}x_i,
\end{equation}
where the bilateral weights are defined as
\begin{equation}
\label{eq:bfweight}
  w_{i,j}=\exp\left(-\frac{(x_i-x_j)^2}{2\sigma_r^2}\right)\exp\left(-\frac{\|\pv_i-\pv_j\|^2}{2\sigma_d^2}\right),
\end{equation}
where $\pv_i$ denotes the coordinate of pixel $i$, and $\sigma_d$ and $\sigma_r$ are smoothing parameters. 
In fact, the bilateral filter can be interpreted as a graph filter \cite{Gadde2013}. In the corresponding graph topology, each node represents a pixel, and is connected to all nodes (including itself) in the $k\times k$ window around it. Let $\Wm$ be the weight matrix with values defined in \eqref{eq:bfweight} and $\Dm$ be its corresponding degree matrix, then the filter output \eqref{eq:bf_output} is $\yv=\Dm^{-1}\Wm\xv$. Define $\hat{\xv}=\Dm^{1/2}\xv$, then we have 
\begin{equation} 
\label{eq:bf_gfilter}
  \hat{\yv}=\Dm^{-1/2}\Wm\Dm^{-1/2}\hat{\xv}=(\Id-\hbox{\boldmath$\Lc$})\hat{\xv}, 
\end{equation}
where $\hbox{\boldmath$\Lc$}=\Dm^{-1/2}\Lm\Dm^{-1/2}$ is the normalized combinatorial Laplacian matrix. Equation \eqref{eq:bf_gfilter} is a graph filter output with spectral response $h(\lambda)=1-\lambda$. 

We consider a variation of this interpretation with a loopless graph. For each node (pixel) $i$ in $\xv$, we consider a new node $i'$ connected to all nodes in the $k\times k$ window centered at $i$, as illustrated in Fig. \ref{fig:bflocal}(b). When the weights are defined as \eqref{eq:bfweight} and we perform a graph filter as in \eqref{eq:bf_gfilter} to the graph of all $i$ and $i'$, $y_{i'}$ will be the bilateral filter output.

%
%
\begin{table}
\caption{Denoising results in PSNR for different graph weights and topologies. N: noisy input. BF: bilateral weights. CGL: closed form solution as in \eqref{eq:cf1}. L: Lena. A: airplane. S: sailboat on lake. P: peppers.}
\label{tab:denoise}
\centering
\begin{tabular}{|c||c||c|c||c|c||c|c||c|c||}
\hline
&& \multicolumn{2}{|c||}{5-neighbor} & \multicolumn{2}{|c||}{$3\times 3$} & \multicolumn{2}{|c||}{$5\times 5$}\\
\cline{2-8} & N & BF & CGL & BF & CGL & BF & CGL \\
\hline
\multirow{4}{*}{L} 
& 15 & 20.59 & {\bf 21.22} & 22.40 & {\bf 23.24} & 24.49 & {\bf 25.56} \\
& 20 & 25.37 & {\bf 25.99} & 26.98 & {\bf 27.73} & 28.40 & {\bf 28.97} \\
& 25 & 29.86 & {\bf 30.46} & 31.11 & {\bf 31.68} & {\bf 31.75} & 31.73 \\
& 30 & 33.98 & {\bf 34.45} & 34.73 & {\bf 34.98} & {\bf 34.83} & 34.17 \\
\hline
\multirow{4}{*}{A} 
& 15 & 20.43 & {\bf 21.04} & 22.10 & {\bf 22.84} & 23.58 & {\bf 24.26} \\
& 20 & 25.00 & {\bf 25.55} & 26.28 & {\bf 26.81} & {\bf 26.92} & 26.88 \\
& 25 & 29.09 & {\bf 29.47} & 29.90 & {\bf 30.08} & {\bf 30.00} & 29.24 \\
& 30 & 32.61 & {\bf 32.63} & {\bf 33.01} & 32.69 & {\bf 32.74} & 31.24 \\
\hline
\multirow{4}{*}{S} 
& 15 & 20.54 & {\bf 21.18} & 22.26 & {\bf 23.06} & 24.01 & {\bf 24.82} \\
& 20 & 25.31 & {\bf 25.96} & 26.79 & {\bf 27.47} & 27.91 & {\bf 28.06} \\
& 25 & 29.83 & {\bf 30.48} & 30.94 & {\bf 31.49} & {\bf 31.55} & 31.16 \\
& 30 & 34.12 & {\bf 34.72} & 35.00 & {\bf 35.13} & {\bf 34.99} & 34.13 \\
\hline
\multirow{4}{*}{P} 
& 15 & 20.57 & {\bf 21.18} & 22.35 & {\bf 23.13} & 24.49 & {\bf 25.44} \\
& 20 & 25.32 & {\bf 25.87} & 26.95 & {\bf 27.60} & 28.52 & {\bf 28.94} \\
& 25 & 29.80 & {\bf 30.22} & 31.13 & {\bf 31.53} & {\bf 32.06} & 31.97 \\
& 30 & 33.66 & {\bf 33.80} & 34.68 & {\bf 36.76} & {\bf 34.85} & 34.30 \\
\hline 
\end{tabular}
\end{table}

Now, we would like to evaluate the alternative weight construction using \eqref{eq:cf1}. Since scaling of the weights does not change the result of \eqref{eq:bf_output}, we use a scaled version of \eqref{eq:cf1}:
\begin{equation}
\label{eq:invms}
  w_{i,j}=\left[1+(x_i-x_j)^2/2\alpha\right]^{-1}.
\end{equation}
In addition to $k\times k$ windows, we also consider a 5-neighbor topology, as shown in Fig. \ref{fig:bflocal_4conn}. For an image with $n$ pixels, the overall graph has $2n$ nodes, and $5n$ or $k^2n$ edges for 5-neighbor topology or $k\times k$ window, respectively. Note that, the overall graph has cycles; however, as discussed previously, \eqref{eq:invms} would be a reasonable approximation for the optimal solution of \eqref{eq:prob_cgl} when the graph is sparse. 

%
%
We apply filtering in \eqref{eq:bf_gfilter} using three different topologies: 5-neighbor, $3\times 3$ window, and $5\times 5$ window. For parameter selection, we use a noise-level-adaptive scheme to achieve robust performance for all noise levels. We choose $\sigma_d=3$ and $\sigma_r=2\sigma_n$, where the linear dependence is recommended in \cite{Liu2006}, and the noise standard deviation $\sigma_n$ is estimated from the noisy image using the method proposed in \cite{Immerkaer1996}. We choose $\alpha=4\sigma_n^2$ accordingly so that $2\sigma_r^2=2\alpha$. From the fact that $\exp(-\gamma)\approx(1+\gamma)^{-1}\approx 1-\gamma$ when $\gamma\approx 0$, \eqref{eq:invms} is close to the first exponential function in \eqref{eq:bfweight} when $x_i-x_j$ is close to 0. We use the grayscale versions of test images, {\it Lena}, {\it airplane}, {\it sailboat on lake}, and {\it peppers}, with Gaussian noise added. 

The results measured in peak-signal-to-noise ratio (PSNR) are shown in Table \ref{tab:denoise}. We observe that the weights given in \eqref{eq:cf1} lead to better denoising results than the conventional bilateral filter when the window size is smaller and when the noise level is higher (a lower noisy PSNR). Indeed, when the parameters are chosen as above, the weights in \eqref{eq:invms} have a slightly stronger smoothing effect than the bilateral weights since $(1+\gamma)^{-1}>e^{-\gamma}$ for $\gamma>0$. Such stronger smoothing filter is more likely to achieve better results for higher noise levels than for lower ones. We also notice that, with a sparser graph topology, the proposed weights yield better results as compared to bilateral weights. Though the 5-neighbor topology does not do as well as $3\times 3$ and $5\times 5$ windows, it has the lowest computational complexity, which could make attractive in practice for certain applications such as video compression.
%
%

\section{Conclusion}
\label{sec:conclusion}
In this work, we focus on the graph learning problem formulated as a MAP parameter estimation for Gaussian Markov random field. In particular, we have investigated sufficient conditions for this problem to have a closed form solution. Then, under acyclic graph topology, we have derived the desired form, which gives rise to a highly efficient tree graph construction from data samples. An error bound when this form is used for a cyclic topology was also provided. We have applied the resulting graph weights to image denoising, to provide an alternative to bilateral weights. Our results show that, for certain window types/sizes and noise levels, the proposed weight construction provides a better denoising result in PSNR on test images than the bilateral filter. 


\bibliographystyle{IEEEbib}
\bibliography{refs}

\vfill\pagebreak

\section{Appendix}
In this appendix we show Theorem \ref{thm:erbd}, together with a slightly tighter bound \eqref{eq:tighter_bound}. Consider a connected\footnote{We always assume the topology to be connected. Otherwise, the weight estimation problem for can be considered as a number of subproblems, each corresponding to a separate connected component of $\Ec$.} cyclic graph topology $\Ec=\{\varepsilon_1,\dots,\varepsilon_m\}$ with $m>n-1$. Let $\Lm^*$ be the optimal solution to \eqref{eq:prob_cgl} and $\Lm$ be the matrix constructed using \eqref{eq:cf1}. We also denote $\uv=(u_1,\dots,u_m)^T$ and $\uv^*=(u_1^*,\dots,u_m^*)^T$ as the weights in $\Lm$ and $\Lm^*$. 
\subsection{Exploiting the Acyclic Optimal Solution}
First we note that the optimal solution $\Lm^*$ corresponds to a connected graph. Although not all weights $u_j^*$ with $j=1,\dots,m$ are necessarily positive, there exists a spanning tree $\Ec_\Tc\subset\Ec$ with $n-1$ edges, such that the weights in $\Lm^*$ associated to edges of this tree are all positive. Take this subtree, and assume without loss of generality $\Ec_\Tc=\{\varepsilon_1,\dots,\varepsilon_{n-1}\}$. Let the set of other edges be $\Ec_\Qc=\Ec\backslash\Ec_\Tc=\{\varepsilon_n,\dots,\varepsilon_m\}$. Then, we can partition the incidence matrix and the weight vectors based on $\Tc$ and $\Qc$: 
\begin{align*}
  \Xim &= \begin{pmatrix} \Xim_\Tc & \Xim_\Qc \end{pmatrix},\quad
  \uv = \begin{pmatrix} \uv_\Tc \\ \uv_\Qc \end{pmatrix},\quad 
  \uv^* = \begin{pmatrix} \uv_\Tc^* \\ \uv_\Qc^* \end{pmatrix}, \\
  \Xim_\Tc &= (\xiv_1,\dots,\xiv_{n-1})\in\mathbb{R}^{n\times (n-1)}, \\
  \Xim_\Qc &= (\xiv_n,\dots,\xiv_m)\in\mathbb{R}^{n\times (m-n+1)}, \\
  \uv_\Tc &= (u_1,\dots,u_{n-1})^T, \quad \uv_\Qc=(u_n,\dots,u_m)^T, \\
  \uv_\Tc^* &= (u_1^*,\dots,u_{n-1}^*)^T, \quad \uv_\Qc^*=(u_n^*,\dots,u_m^*)^T.
\end{align*}

Next, we consider the Laplacian matrices $\Lm$ and $\Lm^*$ supported on $\Ec_\Tc$ only, and denote them as $\Lm_\Tc$ and $\Lm_\Tc^*$:
\begin{align*}
  & \Lm = \Lm_\Tc+\Lm_\Qc, \quad \Lm^* = \Lm_\Tc^*+\Lm_\Qc^*, \\
  & \Lm_\Tc = \Xim_\Tc\;\text{diag}(\uv_\Tc)\;\Xim_\Tc^T,\quad \Lm_\Qc = \Xim_\Qc\;\text{diag}(\uv_\Qc)\;\Xim_\Qc^T, \\
  & \Lm_\Tc^* = \Xim_\Tc\;\text{diag}(\uv_\Tc^*)\;\Xim_\Tc^T,\quad \Lm_\Qc^* = \Xim_\Qc\;\text{diag}(\uv_\Qc^*)\;\Xim_\Qc^T.
\end{align*}
As previous, we denote $\Gm_\Tc=(\Xim,\onev)$, $\uv_\Tc^+=(\uv_\Tc^T,1/n)^T$ and ${\uv_\Tc^*}^+=({\uv_\Tc^*}^T,1/n)^T$. 

From Theorem 1, we know that $\uv_\Tc$ corresponds to the optimal weight solution with $\Ec_\Tc$. That is, $\Jc(\Lm_\Tc)\leq\Jc(\Lm_\Tc^*)$. This gives us a useful inequality: 
\begin{align}
\label{eq:ineq}
  & \Jc(\Lm_\Tc)\leq\Jc(\Lm_\Tc^*) \nonumber\\
  & \Leftrightarrow -\text{logdet}(\Lm_\Tc+\onev\onev^T/n)+\text{trace}(\Lm_\Tc\Km) \nonumber\\
  & \qquad \leq -\text{logdet}(\Lm_\Tc^*+\onev\onev^T/n)+\text{trace}(\Lm_\Tc^*\Km) \nonumber\\
  & \Leftrightarrow \text{trace}((\Lm_\Tc-\Lm_\Tc^*)\Km) \nonumber\\
  & \qquad \leq \text{logdet}(\Lm_\Tc+\onev\onev^T/n)-\text{logdet}(\Lm_\Tc^*+\onev\onev^T/n).
\end{align}
Note that all $u_j$ and $u_j^*$ are positive, so all determinants are not zero. 
We can reduce $\Jc(\Lm)-\Jc(\Lm^*)$ using the above inequality \eqref{eq:ineq}.
\begin{align}
  & \Jc(\Lm)-\Jc(\Lm^*) \nonumber\\
  & = -\text{logdet}(\Lm+\onev\onev^T/n) + \text{logdet}(\Lm^*+\onev\onev^T/n) \nonumber\\
  &\qquad +\text{trace}(\Lm\Km)-\text{trace}(\Lm^*\Km) \nonumber\\
  &= -\text{logdet}(\Lm+\onev\onev^T/n) + \text{logdet}(\Lm^*+\onev\onev^T/n) \nonumber\\
  &\qquad +\text{trace}((\Lm_\Tc-\Lm_\Tc^*)\Km) + \text{trace}((\Lm_\Qc-\Lm_\Qc^*)\Km) \nonumber\\
\label{eq:reduced}
  &\leq \text{trace}((\Lm_\Qc-\Lm_\Qc^*)\Km) \nonumber\\
  &\qquad +\underbrace{\left[\text{logdet}(\Lm^*+\onev\onev^T/n)-\text{logdet}(\Lm_\Tc^*+\onev\onev^T/n)\right]}_{\gamma_1} \nonumber\\
  &\qquad -\underbrace{\left[\text{logdet}(\Lm+\onev\onev^T/n)- \text{logdet}(\Lm_\Tc+\onev\onev^T/n)\right]}_{\gamma_2}.
\end{align}
With $u_j=1/(\xiv_j^T\Km\xiv_j)$, the first term in the right hand side can be simplified as
\begin{align}
  &\text{trace}((\Lm_\Qc-\Lm_\Qc^*)\Km) \nonumber\\
  &= \text{trace}((\text{diag}(\uv_\Qc)-\text{diag}(\uv_\Qc^*))\;\Xim_\Qc^T\Km\Xim_\Qc) \nonumber \\
\label{eq:trace_reduce}
  &= \sum_{j=n}^m (u_j-u_j^*)(\xiv_j^T\Km\xiv_j)
  = \sum_{j=n}^m (u_j-u_j^*)(1/u_j)
  = \sum_{j=n}^m \left(1-\frac{u_j^*}{u_j}\right).
\end{align}

\subsection{Comparison between $\gamma_1$ and $\gamma_2$}
To reduce the other terms in the right hand side of \eqref{eq:reduced}, we write
\begin{align}
\label{eq:L_split}
  \Lm+\onev\onev^T/n
  &= (\Lm_\Tc+\onev\onev^T/n)+\Lm_\Qc \nonumber\\
  &= (\Lm_\Tc+\onev\onev^T/n)+\Xim_\Qc\;\text{diag}(\uv_\Qc)\;\Xim_\Qc^T \\
\label{eq:Lstar_split}
  \Lm^*+\onev\onev^T/n 
  &= (\Lm_\Tc^*+\onev\onev^T/n)+\Lm_\Qc^* \nonumber\\
  &= (\Lm_\Tc^*+\onev\onev^T/n)+\Xim_\Qc\;\text{diag}(\uv_\Qc^*)\;\Xim_\Qc^T.
\end{align}
Then, we use a determinant lemma: for invertible $\Am$ and $\Wm$, 
\begin{equation}
\label{eq:det_lemma}
  \det(\Am+\Um\Wm\Vm^T)=\det(\Am)\det(\Wm)\det(\Wm^{-1}+\Vm^T\Am^{-1}\Um).
\end{equation}
When it is applied to \eqref{eq:L_split}, it gives us 
\begin{align}
  \gamma_2&=\text{logdet}(\Lm+\onev\onev^T/n)-\text{logdet}(\Lm_\Tc+\onev\onev^T/n) \nonumber\\
  &=\text{logdet}(\text{diag}(\uv_\Qc))\nonumber\\
  &\quad +\text{logdet}\left[\text{diag}(\uv_\Qc)^{-1}
  +\Xim_\Qc^T(\Lm_\Tc+\onev\onev^T/n)^{-1}\Xim_\Qc\right]\nonumber\\
\label{eq:gamma2}
  &=\sum_{j=n}^n\log(u_j) \nonumber\\
  &\quad +\text{logdet}\left[\text{diag}(\uv_\Qc)^{-1}+\Xim_\Qc^T\Gm^{-T}(\text{diag}(\uv_\Tc^+))^{-1}\Gm^{-1}\Xim_\Qc\right]
\end{align}
The same lemma cannot be applied directly to \eqref{eq:Lstar_split} because $\uv_\Qc^*$ can have zero elements, making $\Wm$ singular. We will, instead, use an upper bound of $\gamma_1$ such that 1)~it is easy to compare this upper bound with $\gamma_2$ and 2)~it allows us to apply the lemma for a neat reduction. For $j=1,\dots,m$, define weights modified from $u_j^*$ as
\[
  \tilde{u_j}=\left\{\begin{array}{ll} u_j^*, & u_j^*>0 \\ u_j, & u_j^*=0 \end{array}\right.
\]
and let $\tilde{\uv}=(\tilde{u_1},\dots,\tilde{u_m})=(\tilde{\uv_\Tc}^T,\tilde{\uv_\Qc}^T)^T$. Now, it is clear that $\uv_\Qc^*\leq\tilde{\uv_\Qc}$, so we have
\[
  \Lm^*+\onev\onev^T/n \preceq (\Lm_\Tc^*+\onev\onev^T/n)+\Xim_\Qc\;\text{diag}(\tilde{\uv_\Qc})\;\Xim_\Qc^T,
\]
where $\Am\preceq\Bm$ means $\Bm-\Am$ is positive semidefinite, which also implies that $\det(\Am)\leq\det(\Bm)$ and $\det(\Um\Am\Um^T)\leq\det(\Um\Bm\Um^T)$ for any $\Um$.
Applying the determinant lemma \eqref{eq:det_lemma} to this upper bound, we have
\begin{align}
  \gamma_1&=\text{logdet}(\Lm^*+\onev\onev^T/n)-\text{logdet}(\Lm_\Tc^*+\onev\onev^T/n) \nonumber\\
  &\preceq \text{logdet}\left[(\Lm_\Tc^*+\onev\onev^T/n)+\Xim_\Qc\;\text{diag}(\tilde{\uv_\Qc})\;\Xim_\Qc^T\right]\nonumber\\
  &\quad -\text{logdet}(\Lm_\Tc^*+\onev\onev^T/n) \nonumber\\
  &=\text{logdet}(\text{diag}(\tilde{\uv_\Qc})\nonumber\\
  &\quad +\text{logdet}\left[(\text{diag}(\tilde{\uv_\Qc}))^{-1}+\Xim_\Qc^T(\Lm_\Tc^*+\onev\onev^T/n)^{-1}\Xim_\Qc\right] \nonumber\\
\label{eq:gamma1}
  &=\sum_{j=n}^m \log(\tilde{u_j})\nonumber\\
  &\quad+\text{logdet}\left[\text{diag}(\tilde{\uv_\Qc})^{-1}+\Xim_\Qc^T\Gm^{-T}\text{diag}(\tilde{\uv_\Tc}^+)^{-1}\Gm^{-1}\Xim_\Qc\right],
\end{align}
where $\tilde{\uv_\Tc}^+=(\tilde{u_1},\dots,\tilde{u_m},1/n)^T$.

For $j=1,\dots,m$, we define
\[
  r_j = \tilde{u_j}/u_j = \left\{\begin{array}{ll} u_j^*/u_j, & u_j^*>0 \\ 1, & u_j^*=0. \end{array}\right.
\]
Then, we have the following inequalities
\begin{itemize}
\item[i.] Since $r_j/\tilde{u_j}=1/u_j$, we have
\[
  (\text{min}_{j=n}^m r_j)\text{diag}(\tilde{\uv_\Qc})^{-1}\preceq\text{diag}(\uv_\Qc)^{-1}
\]
\item[ii.] For the same reason, we have
\[
  \left(\text{min}_{j=1}^{n-1} r_j\right)\text{diag}(\tilde{\uv_\Tc^+})^{-1}\preceq\text{diag}(\uv_\Tc^+)^{-1}.
\]
The inequality will still hold when we multiply both size by two matrices that are symmetric, so
\begin{align*}
  &\left(\text{min}_{j=1}^{n-1} r_j\right) \Xim_\Qc^T\Gm^{-T} \text{diag}(\tilde{\uv_\Tc^+})^{-1}\Gm^{-1}\Xim_\Qc \nonumber\\
  &\quad\preceq\Xim_\Qc^T\Gm^{-T}\text{diag}(\uv_\Tc^+)^{-1}\Gm^{-1}\Xim_\Qc.
\end{align*}

\item [iii.] Let $\beta=\text{min}\left\{\text{min}_{j=1}^{n-1} r_j,\text{min}_{j=n}^{m} r_j\right\}$, then, from i. and ii.,
\begin{align*}
  &\beta\left[\text{diag}(\tilde{\uv_\Qc})^{-1}+\Xim_\Qc^T\Gm^{-T}\text{diag}(\tilde{\uv_\Tc}^+)^{-1}\Gm^{-1}\Xim_\Qc\right] \nonumber\\
  &\quad\preceq \left[\text{diag}(\uv_\Qc)^{-1}+\Xim_\Qc^T\Gm^{-T}(\text{diag}(\uv_\Tc^+))^{-1}\Gm^{-1}\Xim_\Qc\right].
\end{align*}
Taking determinant of both sides, we obtain
\begin{align}
\label{eq:beta_bound}
  &\beta^{m-n+1}\det\left[\text{diag}(\tilde{\uv_\Qc})^{-1}+\Xim_\Qc^T\Gm^{-T}\text{diag}(\tilde{\uv_\Tc}^+)^{-1}\Gm^{-1}\Xim_\Qc\right] \nonumber\\
  &\quad\preceq \det\left[\text{diag}(\uv_\Qc)^{-1}+\Xim_\Qc^T\Gm^{-T}(\text{diag}(\uv_\Tc^+))^{-1}\Gm^{-1}\Xim_\Qc\right].
\end{align}
\end{itemize}
Plugging \eqref{eq:beta_bound} into \eqref{eq:gamma1} and compare with $\gamma_2$ as in \eqref{eq:gamma2}, we have
\begin{align}
  \gamma_1
  &\leq \sum_{j=n}^m \log(\tilde{u_j})-(m-n+1)\log(\beta)\nonumber\\
  &\quad +\text{logdet}\left[\text{diag}(\uv_\Qc)^{-1}+\Xim_\Qc^T\Gm^{-T}\text{diag}(\uv_\Tc^+)^{-1}\Gm^{-1}\Xim_\Qc\right]\nonumber\\
  &= \sum_{j=n}^m \log(\tilde{u_j})-(m-n+1)\log(\beta)+\left(\gamma_2-\sum_{j=n}^m \log(u_j)\right) \nonumber\\
  &= \gamma_2 + \sum_{j=n}^m \log\left(\frac{\tilde{u_j}}{u_j}\right)-(m-n+1)\log(\beta)
\end{align}
Finally, we have 
\[
  \gamma_1-\gamma_2 \leq \sum_{j=n}^m \log(r_j) -(m-n+1)\log(\beta)
\]
Together with \eqref{eq:trace_reduce}, the tighter bound we have is
\begin{align}
\label{eq:tighter_bound}
  &\Jc(\Lm)-\Jc(\Lm^*) \nonumber\\
  &\quad\leq \sum_{j=n}^m \left(1-\frac{u_j^*}{u_j}\right)+\sum_{j=n}^m \log(r_j) \nonumber\\
  &\qquad-(m-n+1)\log\left(\text{min}\left\{\text{min}_{j=1}^{n-1} r_j,\text{min}_{j=n}^{m} r_j\right\}\right),
\end{align}
where
\[
  r_j = \tilde{u_j}/u_j = \left\{\begin{array}{ll} u_j^*/u_j, & u_j^*>0 \\ 1, & u_j^*=0 \end{array}\right.
\]
We can further reduce \eqref{eq:tighter_bound} using $\min_j r_j$ and $\max_j r_j$ as
\begin{align}
\label{eq:final_bound}
  &\Jc(\Lm)-\Jc(\Lm^*) \nonumber\\
  &\quad\leq \sum_{j=n}^m \left(1-\frac{u_j^*}{u_j}\right)+\sum_{j=n}^m \log\left(\max_j(r_j)\right) \nonumber\\
  &\qquad -(m-n+1)\log\left(\min_j r_j\right) \nonumber\\
  &\quad\leq \sum_{j=n}^m \left(1-\frac{u_j^*}{u_j}\right)+(m-n+1) \log\left(\frac{\text{max}_j(r_j)}{\text{min}_j(r_j)}\right)
\end{align}

\end{document}